\documentclass[a4paper,12pt]{article}

\usepackage{geometry}
\usepackage[english]{babel}

\usepackage{amsmath}
\usepackage{amsfonts}
\usepackage{amstext}
\usepackage{amssymb}
\usepackage{amsthm}
\usepackage{amscd}
\usepackage{mathrsfs}
\usepackage{authblk}

\usepackage{xcolor}
\definecolor{myurlcolor}{rgb}{0,0.5,0}
\definecolor{mycitecolor}{rgb}{0,0,1}
\definecolor{myrefcolor}{rgb}{0.5,0,0}
\usepackage{graphicx}
\usepackage{tikz}
\usepackage{tikz-cd}

\usepackage[pagebackref,draft=false]{hyperref}
\hypersetup{colorlinks,
linkcolor=myrefcolor,
citecolor=mycitecolor,
urlcolor=myurlcolor}

\usepackage[capitalize]{cleveref}
\usepackage{caption}

\newcommand{\gapp}{\mathscr{G}}
\newcommand{\mappa}{\mathscr{M}}

\begin{document}

\title{Groupoids and Coherent states}

\author{F. di Cosmo, A. Ibort}
\affil{\small{Instituto de Ciencias
Matem\'{a}ticas (CSIC - UAM - UC3M - UCM) ICMAT}}
\affil{ \small{Depto. de Matem\'aticas, Univ. Carlos III de Madrid, \\ Avda. de la
Universidad 30, 28911 Legan\'es, Madrid, Spain.}}


\author{G. Marmo}

\affil{\small{Dipartimento di Fisica E. Pancini dell Universit\'a Federico II di Napoli\\
Complesso Universitario di Monte S. Angelo, via Cintia, 80126 Naples, Italy.}
}
\affil{Sezione INFN di Napoli\\
Complesso Universitario di Monte S. Angelo, via Cintia,  80126 Naples, Italy}


\date{}

\maketitle

\begin{abstract}
Schwinger's algebra of selective measurements has a natural interpretation in terms of groupoids.  This approach is pushed forward in this paper to show that the theory of coherent states has a natural setting in the framework of groupoids.  Thus given a quantum mechanical system with associated Hilbert space determined by a representation of a groupoid, it is shown that any invariant subset of the group of invertible elements in the groupoid algebra determines a family of generalized coherent states provided that a completeness condition is satisfied.   The standard coherent states for the harmonic oscillator as well as generalized coherent states for  $f$-oscillators are exemplified in this picture. 
\end{abstract}

\tableofcontents


\section{Introduction} 

\subsection{Sudarshan and the foundations of Quantum Mechanics: The discovery of coherent states}

George Sudarshan   joined the Tata Institute of Fundamental Research
(TIFR) in Bombay  in the spring of 1952 as a research student.
Two years later, Paul A.M. Dirac visited TIFR and gave a course of
lectures on quantum mechanics. George collaborated with K.K. Gupta in the
preparation of the lecture notes. This work gave him the occasion to be in
close contact with Dirac, a formidable opportunity to learn the subject
from the master himself.  This experience shaped George's attitude to
quantum mechanics all his life.  

Later George went to Rochester for his
Ph.D. Thesis.  Immediately after he went to Harvard University  for two years as a
post-doc with J. Schwinger.   From Harvard, George moved back to Rochester
and there in 1963 he conceived a way to describe  quantum states close to
classical beams of light.  He pointed out the importance of using the
coherent states of the photon field in a paper entitled  ``Equivalence of
Semiclassical and Quantum Mechanical Descriptions of Statistical Light
Beams'' (Physical Review Letters, 1963, \textbf{10}(7) 277) \cite{Su63}.

It is with admiration and great affection that we dedicate this paper to
his memory,  confident that he would have liked the idea that  Schwinger's
approach to quantum mechanics and coherent states appear to be intimately
linked.

\subsection{Coherent states, groups and the foundations of Quantum Mechanics}

The discovery of coherent states constituted a leap forward in the understanding of Quantum Mechanics.      In a few years it was clearly realized the connection between coherent states and group theory.   A. Perelomov \cite{Pe72} showed that families of coherent states can be obtained from irreducible representations of Lie groups, deepening the understanding of the relation between group theory and Quantum Mechanics already started by E. Wigner \cite{Wi39} (see also \cite{Ba47} and references therein)  and H. Weyl \cite{We50}.     

Quite recently Ciaglia \textit{et al} suggested \cite{Ci17} that the properties of coherent states could be understood in the broader setting of the `quantizer-dequantizer' formalism put forward by V. Man'ko  and G. Marmo \cite{Ma02} showing that, relevant as it is, the group theoretical background is not strictly necessary to provide a background for generalized coherent states.   

The problem of characterizing `canonical' coherent states remains open though.  That is, given a quantum system, is there any way to assess the existence of a family of generalized coherent states, that is of a family of states large enough to provide complete description of the system and compatible in a natural way with its dynamics? (see more about these notions in the main body of the paper).  

Note that even if later on the construction and properties of coherent states for groups different from the Heisenberg-Weyl group was discussed at length, in Perelomov's paper there is no hint on the dynamical setting of the theory.

This problem is deeply related to a bothering question related to the foundation of Quantum Mechanics: in which way is the abstract Hilbert space used to describe a quantum system related to its dynamics?   

In the many standard presentations of Quantum Mechanics, either the Hilbert space is constructed using some sort of quantization technique or it is postulated as an abstract Hilbert space associated to the quantum system (the latter would be Dirac's favourite \cite{Di81}).   

One way to answer this question would be by considering an algebraic approach and positing that previous to the introduction of a Hilbert space there is a $C^*$-algebra (related to the observables of the system) associated with the system.   Then, any representation $\pi$ of the $C^*$-algebra would provide the desired Hilbert space (by using the GNS construction associated with any state, for instance).  But again, the question remains, how do we determine the $C^*$-algebra associated to the given quantum system?  

An answer to such conundrum would be to consider that such $C^*$-algebra is a primitive notion and must be postulated \textit{ab initio}.   This is reasonable and consistent with R. Haag understanding of the structure of local quantum systems \cite{Ha96}.   However, it would be desirable to be able to build, or at  least to `guess', the structure of such algebra from the specific experiments  performed on our test system.    

There is a natural way to do that and it relies on J. Schwinger's algebra of selective measurements.  The most relevant feature of Schwinger's description of quantum mechanical systems is that it leads naturally to the notion of groupoid as the basic mathematical structure beneath the description of quantum systems.

\subsection{Schwinger's algebra of measurements and the groupoid picture of Quantum Mechanics}

\subsubsection{Schwinger's algebra of selective measurements}

J. Schwinger's algebra of selective measurements is a  ``\textit{mathematical language that constitutes a symbolic expression of the properties of microscopic measurements}'' \cite{Sc70}.   In Schwinger's algebraic depiction of a quantum mechanical system\footnote{Throughout this paper we will be referring to the original edition of Schwinger's book whose notation we keep using as it is closer to the spirit of this work.}, we will denote by $\mathbf{A}$ a family of compatible physical quantitites and by $\mathbf{a}$ the outcomes obtained by measuring the quantities $\mathbf{A}$ on the given system.  We will call a selective measurement, and we denote it by $M(\mathbf{a}',\mathbf{a})$, the procedure that rejects all systems of an ensemble whose outcomes are different from $\mathbf{a}$ and those accepted are changed in such a way that their outcomes are $\mathbf{a}'$.  The state of such a quantum system is established by performing on it a complete selective measurement (see \cite[Chap. 1]{Sc70}).  Then it is obvious that the family of selective measurements $M(\mathbf{a}',\mathbf{a})$ satisfy the obvious relations:
$$
M(\mathbf{a}'',\mathbf{a}') \circ M(\mathbf{a}',\mathbf{a}) = M(\mathbf{a}'',\mathbf{a}) \, ,
$$
where the natural composition law of selective measurements $M(\mathbf{a}'',\mathbf{a}') \circ M(\mathbf{a}',\mathbf{a})$ is defined as the selective measurement obtained by performing first the selective measurement $M(\mathbf{a}',\mathbf{a})$ and immediately afterwards the selective measurement $M(\mathbf{a}'',\mathbf{a}')$.  If we denote by $ M(\mathbf{a})$ the selective measurement $M(\mathbf{a},\mathbf{a})$, that is the process that filters the systems whose outcomes are $\mathbf{a}$ without changing them, then:

\begin{equation}\label{schwinger_units}
 M(\mathbf{a}')\circ M(\mathbf{a}',\mathbf{a}) = M(\mathbf{a}',\mathbf{a}) \, , \qquad  M(\mathbf{a}',\mathbf{a}) \circ  M(\mathbf{a})= M(\mathbf{a}',\mathbf{a}) \, ,
\end{equation}
It is clear that performing two selective measurements $M(\mathbf{a}',\mathbf{a})$, and  $M(\mathbf{a}''',\mathbf{a}'')$ one after the other will produce a selective measurement again only if $\mathbf{a}'' = \mathbf{a}'$. Otherwise, if $\mathbf{a}'' \neq \mathbf{a}'$, then  $ M(\mathbf{a}''',\mathbf{a}'') \circ M(\mathbf{a}',\mathbf{a}) = \emptyset$ which is not a selective measurement of the form $ M(\mathbf{a}',\mathbf{a})$. 

Notice that if we have three selective measurements $M(\mathbf{a},\mathbf{a}')$,  $M(\mathbf{a}',\mathbf{a}'')$ and $M(\mathbf{a}'',\mathbf{a}''')$ then, because of the basic definitions, the associativity of the composition law holds:
\begin{equation}\label{schwinger_associative}
M(\mathbf{a},\mathbf{a}') \circ (M(\mathbf{a}',\mathbf{a}'') \circ M(\mathbf{a}'',\mathbf{a}''')) = (M(\mathbf{a},\mathbf{a}') \circ M(\mathbf{a}',\mathbf{a}'')) \circ M(\mathbf{a}'',\mathbf{a}''') \, .
\end{equation}
Finally it is worth to observe that given a measurement symbol $M(\mathbf{a}',\mathbf{a})$ the measurement symbol $M(\mathbf{a},\mathbf{a}')$ satisfies:
\begin{equation}\label{schwinger_inverse}
M(\mathbf{a}',\mathbf{a}) \circ M(\mathbf{a},\mathbf{a}') = M(\mathbf{a}') \, ,\quad M(\mathbf{a},\mathbf{a}') \circ M(\mathbf{a}',\mathbf{a}) = M(\mathbf{a}) \, .
\end{equation}

We conclude that the composition law of selective measurements determines a groupoid law in the collection $\mathbf{G}_\mathbf{A}$ of all measurement symbols $M(\mathbf{a}',\mathbf{a})$ associated with the complete family of observables $\mathbf{A}$, whose objects (the events of the system) are the possible outcomes $\mathbf{a}$ of the observables $\mathbf{A}$.

\subsubsection{The groupoid picture of Quantum Mechanics}

Schwinger's observation has deep implications. It shows that the algebraic structure of the quantities describing quantum systems is that of the algebra of a groupoid such groupoid determined by the family of physical transitions between possible outcomes of the system \cite{Ci17}.   

Thus we will assume in what follows that a quantum system is described starting with a groupoid $\mathbf{G}$ whose elements $\alpha$ correspond to physical transitions of the system (in Schwinger's conceptualization the `transitions'  $\alpha$ are the selective measurements $M(\mathbf{a}', \mathbf{a})$ described before) and whose objects $x$ represent possible outcomes of physical magnitudes (again, in Schwinger's description, the outcomes $x$ correspond to actual measurements of physical observables).   
Thus a groupoid provides the kinematical background for the description of a quantum system (as we have not yet introduced a dynamics into the picture, see \cite{Ci17},\cite{Ib18a}, \cite{Ib18b}  for more details).  Hence in what follows we will succinctly review the main notions on groupoids that will be needed later.

The abstract definition of a groupoid $\mathbf{G}$ is that of a category all whose morphisms are invertible.    Because the aim of this paper is to present the main ideas towards a theory of coherent states on quantum mechanical systems constructed out of first principles we assume for the rest of the paper that we will be dealing with concrete groupoids. For the sake of simplicity and clarity, in describing the basic notions related to groupoids, we will mainly refer to finite groupoids, where intuition can be helpful for the understanding of many properties. However, many of the results presented in what follows can be extended to larger classes of groupoids, as it will be discussed elsewhere. Moreover, it will be assumed that groupoids are small categories, hence a set theoretical notation for families of morphisms and objects will be used without further mention (see \cite{Ib19b} for further details).

Objects in the groupoid $\mathbf{G}$ will be denoted $x,y$, etc., and its morphisms by greek letters $\alpha, \beta$, etc., adopting a convenient diagrammatic notation $\alpha \colon x \to y$, where $x$ is the source of the morphism $\alpha$ and $y$ its target.   The family of objects of $\mathbf{G}$ will be denoted by $\Omega$ and the family of morphisms will be denoted simply by $\mathbf{G}$ again.   The source and target maps sending any morphism into its source and target will be denoted by $s \colon \mathbf{G} \to \Omega$ and $t \colon \mathbf{G} \to \Omega$ respectively.  Thus $s(\alpha) = x$, $t(\alpha) = y$ indicate that $\alpha \colon x \to y$.    A groupoid $\mathbf{G}$ over the space of objects $\Omega$ will be sometimes denoted as $\mathbf{G}  \rightrightarrows  \Omega$ to emphasize the source and the target maps $s,t$.   

In this paper the set of morphisms from $x$ to $y$ will be denoted by $\mathbf{G}(y,x)$ (while the standard notation in category theory would be Hom$(x,y)$)\footnote{Notice the backwards notation for the source and the target of morphisms $\alpha \colon x \to y$ in the sets $\mathbf{G}(y,x)$.}.   It is clear that $\mathbf{G}(y,x) = s^{-1}(x) \cap t^{-1}(y)$.   If $\mathbf{G}$ is finite, the number of morphisms of $\mathbf{G}$ will be called the order of $\mathbf{G}$ and denoted by $| \mathbf{G} |$.  
Similarly the number of objects will be denoted by $| \Omega |$.   

The composition law of the groupoid will be denoted by $\circ$ and the morphisms $\alpha$ and $\beta$ will be said to be composable if $t(\alpha) = s(\beta)$ in which case the composition will be denoted $\beta \circ \alpha$ (again, notice the backwards convention for the composition). Thus $\circ \colon \mathbf{G}(z,y) \times  \mathbf{G}(y,x) \to \mathbf{G}(z,x)$, $\beta \circ \alpha \colon x \to z$ if $\alpha \colon x \to y$ and $\beta \colon y \to z$, and $ \mathbf{G}(z,y) \circ \mathbf{G}(y,x) \subset \mathbf{G}(z,x)$.  The set of composable morphisms will be denoted by $\mathbf{G}_2$, that is:   $\mathbf{G}_2 = \{ (\alpha, \beta)\in \mathbf{G}\times \mathbf{G} \mid t(\alpha) = s(\beta) \}$.

The composition law $\circ$ is associative, that is:
$$
\gamma \circ (\beta \circ \alpha )= (\gamma \circ \beta) \circ \alpha \, , 
$$
whenever the composition of $\alpha$, $\beta$ and $\gamma$ makes sense.

 The unit morphisms of the groupoid will be denoted by $1_x$, $x \in \Omega$, and they satisfy $1_y \circ \alpha = \alpha$ and $\alpha \circ 1_x = \alpha$ for all $\alpha \colon x \to y$.   The family of units $1_x$ defines a canonical inclusion map $i \colon \Omega \to \mathbf{G}$, $i(x) = 1_x$, such that $s \circ i = t \circ i = \mathrm{id}_\Omega$. 
 Any morphism $\alpha \colon x \to y$ is invertible, its inverse will be denoted by $\alpha^{-1}$ and $\alpha^{-1} \colon y \to x$ satisfies $\alpha^{-1} \circ \alpha = 1_x$ and $\alpha \circ \alpha^{-1} = 1_y$.  
 
Given an object $x$ of the groupoid $\mathbf{G}$, the family of all morphisms $\alpha \colon x \to x$ with source and target  the element $x$ in $\Omega$, form a group called the isotropy group at $x$ and denoted by $G_x$, that is, $G_x = \mathbf{G}(x,x)$.
We will denote by $\mathbf{G}_+ (x)$ the family of all morphisms whose source is $x$.  Similarly $\mathbf{G}_-(x)$ is the family of all morphisms whose target is $x$. Then  $\mathbf{G}(y,x) = \mathbf{G}_+ (x) \cap \mathbf{G}_-(y)$, and  $G_x = \mathbf{G}_+ (x) \cap \mathbf{G}_-(x)$.    Notice that if $\mathbf{G}(y,x)$ is non-void, then the isotropy groups $G_y$ and $G_x$ are isomorphic (it suffices to check that if $\alpha \colon x \to y$, then the map $\varphi_\alpha \colon G_x \to G_y$ given by $\phi_\alpha (\gamma_x) = \alpha \circ \gamma_x \circ \alpha^{-1}$ is a group isomorphism). In particular, for finite groupoids $|\mathbf{G}(y,x) | = |G_x| = | G_y|$.

Note that groups $G$ are particular instances of groupoids with the peculiarity that their object space $\Omega$ consists on a single element, usually identified with the neutral element $e$ of the group. 
 
A subgroupoid $\mathbf{H}$ of the groupoid $\mathbf{G}$ is a groupoid which is a subcategory of $\mathbf{G}$.  If $\mathbf{G}$ is a finite groupoid, a subgroupoid $\mathbf{H}$ must be finite too and then, the functor $j \colon \mathbf{H} \to \mathbf{G}$ that describes $\mathbf{H}$ as a subcategory defines an injective map (denoted with the same symbol) $j \colon \mathbf{H} \to \mathbf{G}$ that maps morphisms in $\mathbf{H}$ into morphisms in $\mathbf{G}$, $\alpha' \to \alpha = j(\alpha')$. The functor $j$ defines also an injective map (again denoted with the same symbol) between the objects $\Omega'$ of $\mathbf{H}$ and the objects $\Omega$ of $\mathbf{G}$, that is $j(x') = x$.   Moreover $j(1_{x'}) = 1_{j(x')} = 1_x$.   The map $j$ satisfies the obvious compatibility condition: $s \circ j = j \circ s'$, $t \circ j = j \circ t'$, where $s'$ and $t'$ denote the source and target maps of $\mathbf{H}$ respectively.     

Given a subgroupoid $\mathbf{H}$ of $\mathbf{G}\rightrightarrows \Omega$ with space of objects $\Omega'$ smaller than  $\Omega$, we may always consider another subgroupoid $\widetilde{\mathbf{H}}$, that extends $\mathbf{H}$ naturally as a subgroupoid of $\mathbf{G}$, whose space of objets is $\Omega$;  $\widetilde{\mathbf{H}}$ is defined by simply adding the units $1_x$, $x \in \Omega \backslash \Omega'$ to $\mathbf{H}$.
Hence, in what follows a subgroupoid $\mathbf{H}$ of the finite groupoid $\mathbf{G}$ will be assumed to have the same space of objects as $\mathbf{G}$; it will be identified with the subset $j(\mathbf{H}) \subset \mathbf{G}$ and its morphisms $\alpha'$ will be identified with the corresponding morphisms $\alpha = j(\alpha')$ of $\mathbf{G}$.

Given two finite groupoids $\mathbf{G}_a$, with object spaces $\Omega_a$, $a = 1,2$, we define its coproduct (or disjoint union) as the groupoid, denoted by $\mathbf{G}_1 \sqcup \mathbf{G}_2$,  whose morphisms are the disjoint union of the morphisms in $\mathbf{G}_1$ and $\mathbf{G}_2$ and whose objects are the disjoint union of the objects $\Omega_1$ and $\Omega_2$.  The composition law and source and target maps are the obvious ones.  It is clear that both $\mathbf{G}_1$ and $\mathbf{G}_2$ are subgroupoids of $\mathbf{G}_1 \sqcup \mathbf{G}_2$ with the canonical inclusion functors $j_a \colon \mathbf{G}_a \to \mathbf{G}_1 \sqcup \mathbf{G}_2$, $a = 1,2$.

Given an object $x \in \Omega$, the orbit  $\mathcal{O}_x$ of the groupoid $\mathbf{G}$ through $x$ is the collection of objects corresponding to the targets of morphisms in $\mathbf{G}_+(x)$, that is, $y \in \mathcal{O}_x$ if there exists $\alpha \colon x \to y$ or, in other words, $\mathcal{O}_x = t(\mathbf{G}_+(x))$.  We will say that the groupoid $\mathbf{G}$ is connected (or transitive) if it has just one orbit, i.e. $\mathbf{G}$ is connected if for any $x,y$ objects, there is a morphism $\alpha \colon x \to y$.   Notice that the isotropy groups $G_x$, $G_y$ corresponding to objects $x,y$ in the same orbit are isomorphic (even if not canonically isomorphic).  

Let $\mathbf{G}$ be a finite groupoid over the space of objects $\Omega$.  Let us denote by $\Omega /\mathbf{G}$ the space of orbits $\mathcal{O}_x$ of $\Omega$.   The orbits $\mathcal{O} \in \Omega /\mathbf{G}$ define a partition of $\Omega$.  We will denote by $\mathbf{G}_\mathcal{O}$ the restriction of the groupoid $\mathbf{G}$ to the orbit $\mathcal{O}$, that is $\alpha \colon x \to y \in \mathbf{G}_\mathcal{O}$ if $x,y \in \mathcal{O}$. The groupoid $\mathbf{G}_\mathcal{O}$ is a subgroupoid of $\mathbf{G}$ and is a connected groupoid over $\mathcal{O}$.  Then the groupoid $\mathbf{G}$ is the direct union of the connected groupoids $\mathbf{G}_\mathcal{O}$:  
\begin{equation}\label{partition}
\mathbf{G} = \bigsqcup_{\mathcal{O}\in \Omega/\mathbf{G}} \mathbf{G}_\mathcal{O} \, .
\end{equation}
Thus any groupoid $\mathbf{G}$  is a disjoint union of connected groupoids and its structure will be described by determining the structure of the corresponding connected subgroupoids.   


\subsubsection{Groupoids and representations}\label{sec:representations}

An interesting feature of the groupoids picture of quantum mechanics is that they provide a family of natural  representations with support Hilbert spaces that may be considered as naturally associated with the system.    

A unitary representation $U$ of the groupoid $\mathbf{G} \rightrightarrows \Omega$ is a family of Hilbert spaces $\mathcal{H}_x$, $x \in \Omega$ and a family of unitary maps $U(\alpha) \colon \mathcal{H}_x \to \mathcal{H}_y$, $\alpha \colon x \to y \in \mathbf{G}$, such that:
$$
U(1_x) = \mathrm{id \, }_{\mathcal{H}_x} \, , \quad U(\beta \circ \alpha ) = U(\beta) U(\alpha) \, , (t(\alpha )= s(\beta) )\, , \quad U(\alpha^{-1}) = U(\alpha)^\dagger \, .
$$
The representation $U$ will be said to be locally finite if the Hilbert spaces $\mathcal{H}_x$ are finite dimensional.    Note that if the groupoid $\mathbf{G}$ is a group, that is the object space $\Omega$ consists of just one element, then the notion of unitary representation above becomes that of a unitary representation of a group.

We can form the Hilbert space $\mathcal{H} = \bigoplus_{x \in \Omega} \mathcal{H}_x$, that will be called the total support space of the representation $U$ and it will be often denoted by $(U,\mathcal{H})$.   

Given a groupoid $\mathbf{G} \rightrightarrows \Omega$, we may define its groupoid algebra $\mathbb{C}[\mathbf{G}]$ as the associative algebra generated by the morphisms $\alpha$ of the groupoid.  Thus elements in $\mathbb{C}[\mathbf{G}]$ will be finite formal linear combinations $a = \sum_{\alpha \in \mathbf{G}} a_\alpha \alpha$, $a_\alpha \in \mathbb{C}$.   Note that if the groupoid is finite the algebra $\mathbb{C}[\mathbf{G}]$ is finite dimensional with dimension $|\mathbf{G}|$.   The natural involution $a \mapsto a^*$ is defined as $a^* = \sum_\alpha \bar{a}_\alpha \alpha^{-1}$ making $\mathbb{C}[\mathbf{G}]$ into a $*$-algebra.

The most conspicuous representation of a groupoid is the fundamental representation $\pi_0$, in which case $\mathcal{H}_x = \mathbb{C}$ for all $x \in \Omega$, and $\pi_0(\alpha)|x \rangle = | y \rangle$, $\alpha \colon x \to y$, with $|x\rangle$, $|y \rangle$ denoting the abstract unitary vectors on the linear spaces $\mathcal{H}_x$ and $\mathcal{H}_y$ respectively.

If the groupoid  is finite the total support Hilbert space is just $\mathcal{H} = \bigoplus_{x \in \Omega} \mathcal{H}_x$. In the particular instance of the fundamental representation we will denote the support Hilbert space as $\mathcal{H}_\Omega$.  Then $\mathcal{H}_\Omega = \bigoplus_{x \in \Omega} \mathbb{C} |x \rangle$. 

We may use the fundamental representation $\pi_0$ to define a norm on the groupoid algebra $\mathbb{C}[\mathbf{G}]$ as follows: $|| a || = || \pi_0(a) ||_\Omega$, and with $|| \cdot ||_\Omega$ the operator norm in $\mathcal{H}$.    It is a trivial exercise to check that $|| a^* a || = ||a ||^2$, thus making $\mathbb{C}[\mathbf{G}]$ into a $C^*$-algebras.

Given a unitary representation $(U, \mathcal{H})$ of the groupoid $\mathbf{G}$, there is a natural $\mathbb{C}[\mathbf{G}]$-module structure on $\mathcal{H}$ given by $a \cdot |\psi \rangle = \sum_\alpha a_\alpha U(\alpha) |\psi \rangle$.  Conversely, given a $\mathbb{C}[\mathbf{G}]$-module structure on the Hilbert space $\mathcal{H}$ there is a unitary representation $U$ of $\mathbf{G}$ defined as $U(\alpha) |\psi_x\rangle = \alpha \cdot |\psi_x\rangle$ with $|\psi_x \rangle \in \mathcal{H}_x = P_x(\mathcal{H})$ and $P_x$ is the orthogonal projector defined by multiplication by $1_x$. Clearly there is a one-to-one correspondence between unitary representations $(U, \mathcal{H})$ of the finite groupoid $\mathbf{G}$ and $\mathbb{C}[\mathbf{G}]$-module structures on the Hilbert space $\mathcal{H}$.  Note that alternatively, a $\mathbb{C}[\mathbf{G}]$-module structure on the Hilbert space $\mathcal{H}$ can be thought of as a representation of the $C^*$-algebra $\mathbb{C}[\mathbf{G}]$, i.e., as a map $\pi \colon \mathbb{C}[\mathbf{G}] \to \mathrm{End\,}(\mathcal{H})$, given by $\pi (a) |\psi \rangle = a \cdot |\psi \rangle = \sum_\alpha a_\alpha U(\alpha ) |\psi_x \rangle$ (see \cite{Ib19} for a recent exposition of the theory of representations of finite groupoids).

Previous notions can be extended without pain to discrete countable groupoids.  In such case we will rather use, instead of the fundamental representation $\pi_0$, the regular representation to define an associated $C^*$-algebra of the groupoid $\mathbf{G}$. This representation will be denoted by $\lambda \colon \mathbb{C}[\mathbf{G}] \to \mathrm{End\,}(\mathcal{H})$, with $\mathcal{H} = l^2(\mathbf{G})$ as support space, and $(\lambda (\alpha) \psi ) (\beta) = \psi (\alpha^{-1}\circ \beta)$, provided that $t(\alpha ) =t(\beta)$ and zero otherwise. Now consider the von Neumann algebra generated by the family of operators $\lambda (a) $, $a \in \mathbb{C}[\mathbf{G}]$, i.e., the double commutant of $\lambda(\mathbb{C}[\mathbf{G}])$. We will denote such algebra by $C^*(\mathbf{G})$ and we will call it the (reduced) $C^*$-algebra of the groupoid $\mathbf{G}$.   

 If the groupoid $\mathbf{G}$ is not discrete but it is a topological groupoid with a locally compact topology, then we may define its associated $C^*$-algebra in a similar way as before by using an auxiliary system of Haar measures (see \cite{Re80} and \cite{La98}).  
We will illustrate the previous notions by discussing a fundamental example.

\subsection{A simple example: the harmonic oscillator}\label{harmonic oscillator}

We will discuss now the paradigmatic example of the harmonic oscillator and its standard coherent states from the perspective of groupoids.  

\subsubsection{The groupoid $\mathbf{G}(A_\infty)$}\label{sec:A_infty}

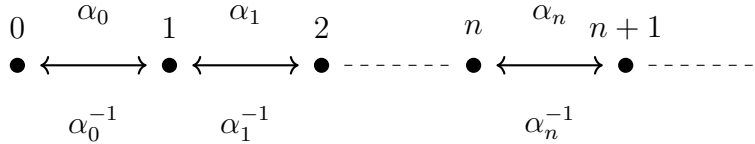
\begin{figure}[h]
\centering
\begin{tikzpicture} 
\fill (0,0) circle  (0.1);
\fill (2,0) circle  (0.1);
\draw [thick,<->] (0.3,0) -- (1.7,0);
\node at (0,0.5)  {$0$};
\node at (2,0.5)  {$1$};
\node [above] at (1,0.4)   {$\alpha_0$};
\node [below] at (1,-0.4)   {$\alpha_0^{-1}$};
\fill (4,0) circle  (0.1);
\draw [thick,<->] (2.3,0) -- (3.7,0);
\node at (4,0.5)   {$2$};
\node [above] at (3,0.4)   {$\alpha_1$};
\node [below] at (3,-0.4)   {$\alpha_1^{-1}$};
\fill (6,0) circle  (0.1);
\draw [dashed] (4.3,0) -- (5.7,0);
\node at (6,0.5)   {$n$};
\fill (8,0) circle  (0.1);
\draw [thick,<->] (6.3,0) -- (7.7,0);
\node at (8,0.5)   {$n+1$};
\node [above] at (7,0.4)   {$\alpha_n$};
\node [below] at (7,-0.4)   {$\alpha_n^{-1}$};
\draw [dashed] (8.3,0) -- (9.7,0);
\end{tikzpicture}
\caption{The diagram $K_\infty$ generating the quantum harmonic oscillator.}
\label{infinite}
\end{figure}

The kinematical description of the harmonic oscillator fits inside a family of system whose physical outcomes and transitions are described by the graph $A_\infty$, that is, the outcomes are labelled by symbols $a_n$, $n= 0,1,2,$..., and the groupoid structure is generated by the family of transitions $\alpha_n \colon a_n \to a_{n+1}$ for all $n$ (see Fig. \ref{infinite}).   

The assignment of physical meaning to the outcomes $a_n$ and the transitions $\alpha_n$, that is, their identification with outcomes of a certain observable $A$ and certain physical transitions corresponding to such outcomes will depend on the specific system under study.   

As a particular instance we may consider that the outcomes are identified with the energy levels of a given system (the spectrum of the Hamiltonian), an atom, or the number of photons of a given frequency in a cavity for example.  In the case of atoms the transitions will correspond to the physical transitions observed by measuring the photons emitted or absorbed by the system.  In the case of an e.m. field in a cavity, the transitions will correspond to the change in the number of photons that could be determined by counting the photons emitted by the cavity or pumping a determined number of photons into it.   

At this point, no specific values have been assigned to the events $a_n$ and transitions $\alpha_n$, they just represent the kinematical background for the theory.     An assignment of numerical values to them will correspond to determining the dynamical prescription of the system.  For instance, in the case of energy levels, we will be assigning a real number $E_n$ to each event while in the case of photons, it will be a certain collection of non-negative integers $n_1,n_2,...$.     In what follows we will focus on the simplest non-trivial assignment of the number $n$ to the event $a_n$.

The groupoid of transitions $\mathbf{G}(A_\infty)$ generated by this system is the groupoid of pairs of natural numbers or, in other words, the complete graph with countable many vertices $K_\infty$, labelled by non-negative integers, $n = 0,1,2,\ldots,$ and they constitute its space of objects $\Omega = K_\infty$.    Transitions $m \to n$ will be denoted by $\alpha_{n,m}$ or just $(n,m)$ for short.  The notation in the picture (Fig. \ref{infinite}) corresponds to $\alpha_n := \alpha_{n+1,n} = (n+1,n)$.
With this notation, two transitions  $(n,m)$ and $(j,k)$ are composable if and only if $m = j$, and their composition will be given by $(n,m)\circ (m,k) = (n,k)$. Notice that $(n,m)^{-1} = (m,n)$ and $1_n = (n,n)$ for all $n \in \mathbb{N}$. Note that as a set the groupoid $\mathbf{G}(A_\infty)$ is just the Cartesian product $\mathbb{N} \times \mathbb{N}$. In what follows we will just denote the groupoid $\mathbf{G}(A_\infty)$ as $\mathbf{A}_\infty$ for brevity.

The algebra of the groupoid $\mathbf{A}_\infty$ could be described as a $C^*$-algebra of functions on the groupoid $\mathbf{A}_\infty$.  To construct it we start by considering the set of functions which are zero except for a finite number of transitions, i.e., of pairs $(n,m)$, and then we will take the closure with respect to an appropriate topology.  Thus, denote by $\mathcal{F}_\mathrm{alg} (\mathbf{A}_\infty)$  the set of functions on $\mathbf{A}_\infty$ which are zero except for a finite number of pairs $(n,m)$.   We may write any one of these functions as:
\begin{equation}\label{expansion_inf}
f = \sum_{n,m= 1}^\infty f(n,m) \delta_{(n,m)} \, ,
\end{equation}
where only a finite number of coefficients $f(n,m)$ are different from zero.  The function $\delta_{(n,m)}$ is the obvious delta function $\delta_{(n,m)}(\alpha_{jk}) = \delta_{(n,m)} (j,k) =  \delta_{nj}\delta_{mk}$.    

The involution $f \mapsto f^*$ in the algebra $\mathcal{F}_\mathrm{alg} (\mathbf{A}_\infty)$ is defined in the standard way $f^*(n,m) = \overline{f(m,n)}$ for all $n,m$.
Note that we may interpret functions $f$ in $\mathcal{F}_\mathrm{alg} (\mathbf{A}_\infty)$ as a formal linear combinations of elements $(n,m) \in \mathbf{A}_\infty$, that is we can identify $\mathcal{F}_\mathrm{alg} (\mathbf{A}_\infty)$ with the groupoid algebra $\mathbb{C}[\mathbf{A}_\infty]$ discussed before.

Given two functions $f, g \in \mathcal{F}_{alg} (\mathbf{A}_\infty )$ we define its convolution product $f\star g$ as the function on $\mathcal{F}_{alg} (\mathbf{A}_\infty )$ whose $(n,m)$ coefficient is given by:
$$
(f \star g) (n,m)  = \sum_{(n,j) \circ (j,m) = (n,m)} f(n,j) g(j,m) =  \sum_{j} f(n,j) g(j,m) \, .
$$
Note that $\delta_{(n,m)}\star \delta_{(j,k)} = \delta_{mj} \delta_{(n,k)}$.  Moreover $(f \star g)^\ast = g^\ast \star f^\ast$.
Hence, using Heisenberg's interpretation of observables as (infinite) matrices, we may consider the coefficients $f(n,m)$, $n, m = 0,1,\ldots,$ in the expansion (\ref{expansion_inf}) as defining an infinite matrix $F$ whose entries $F_{nm}$ are the numbers $f(n,m)$. In doing so the convolution product on the algebra $\mathcal{F}_{alg} (\mathbf{A}_\infty )$ becomes the matrix product of the matrices $F$ and $G$ corresponding to $f$ and $g$ respectively (notice that the product is well defined as there are only finitely many non zero entries on both matrices). 

The fundamental representation $\pi_0$ of the system will be supported on the Hilbert space $\mathcal{H}_\Omega$ generated by the vectors $|n\rangle$, $n = 0,1,\ldots$, that is, the family of vectors $\{ |n\rangle \mid  n \in \mathbb{N} \}$ defines an orthonormal basis of $\mathcal{H}_\Omega$.  Thus, the Hilbert space $\mathcal{H}_\Omega$ can be identified with the Hilbert space $l^2(\mathbb{Z})$ of infinite sequences $z = (z_0,z_1,z_2, \ldots )$ of complex numbers with $|| z ||^2 = \sum_{n = 0}^\infty |z_n|^2 < \infty$.  The fundamental representation $\pi_0$ is just given by (recall the definition or the fundamental representation in Sect. \ref{sec:representations}):
$$
\pi_0 (\alpha_{nm}) |k\rangle = \delta_{mk} |n\rangle \, ,
$$
that is, $\pi_0 (\alpha_{nm})$ is the operator in $\mathcal{H}$ that maps the vector $|m\rangle$ into the vector $|n\rangle$ and zero otherwise or, using Dirac's notation $\pi_0 (\alpha_{nm}) = | n \rangle \langle m |$.     

We may use the fundamental representation $\pi_0$ to define a norm on $\mathcal{F}_{alg} (\mathbf{A}_\infty )$ as in Sect. \ref{sec:representations}, that is: $|| f || = || \pi_0(f) ||_{\mathcal{H}}$, and consider its completion with respect to it.   It is clear that such completion is a $C^*$-algebra because:
 $$
 || f^*\star f || = || \pi(f^* \star f) ||_{\mathcal{H}} = || \pi_0(f^*)\pi (f) ||_{\mathcal{H}} = || \pi_0(f)^\dagger\pi_0 (f) ||_{\mathcal{H}} = || \pi_0(f) ||^2_{\mathcal{H}} = || f ||^2 \, .
 $$   
 Moreover, by construction, the representation $\pi_0$ is continuous and has a continuous extension to the completed algebra $\overline{\mathcal{F}_{alg} (\mathbf{A}_\infty )}$.  By construction the map $\pi_0$ defines an isomorphism of algebras between the algebra  $\overline{\mathcal{F}_{alg} (\mathbf{A}_\infty )}$ and the algebra $\mathcal{K}(\mathcal{H})$ of compact operators on the Hilbert space $\mathcal{H}$, because compact operators are the closure in the operator norm of the subalgebra of finite rank operators.  Unfortunately the $C^*$-algebra $\mathcal{K}(\mathcal{H})$ is too small for the purposes of describing the dynamics of a quantum system.    
 
 Then it is more convenient to proceed as suggested at the end of the previous section and consider the regular representation instead.
However in the case of groupoids of pairs, this is not strictly necessary as it is possible to show that the regular representation is nothing else that the fundamental representation (with the appropriated multiplicity, see for instance \cite{Ib19b}).    The identification proceeds as follows.  Let $\mathbf{G}(\Omega)$ be the groupoid of pairs of the set $\Omega$, i.e., its transitions are pairs $(y,x)$. Then select an outcome $x \in \Omega$ and consider the subset $\mathbf{G}_+ (x)$ of transitions starting from $x$ and the corresponding space of functions $W_x^+ = \mathcal{F}(\mathbf{G}_+ (x))$.  It is clear that the left regular representation $\lambda$ leaves this subspace invariant (because we are composing on the left thus leaving the space of transitions starting at $x$ invariant).  Moreover we can define a one-to-one correspondence between function in $W_x^+$ and functions in $\Omega$ by $\psi (y) = \Psi (y,x)
$ where $\Psi \in W_x^+$ and $\psi \in \mathcal{F}(\Omega)$.  This correspondence provides the identification between the left regular representation $\lambda$ (restricted to the subspace $W_x^+$) and the fundamental representation $\pi_0$.

Then we may consider the algebra of operators $\pi_0(\mathcal{F}_{alg}(\mathbf{A}_\infty))$ as a subalgebra of the algebra $\mathcal{B}(H_{\Omega})$ of bounded operators on $\mathcal{H}_\Omega$ and the von Neumann algebra generated by it, that is, its double commutant (or the closure on the weak operator topology, or even in the strong operator topology).  It is not hard to check that it coincides with the full algebra of bounded operators on $\mathcal{H}$ because clearly, the commutatnt $\pi_0(\mathcal{F}_{alg}(\mathbf{A}_\infty))'$ consists just in multiples of the identity and, consequently $\pi_0(\mathcal{F}_{alg}(\mathbf{A}_\infty))'' = \mathcal{B}(\mathcal{H}_\Omega)$.  Then we conclude that the $C^*$-algebra $C^*(\mathbf{A}_\infty)$ associated to the groupoid $\mathbf{A}_\infty$ is just the unital $C^*$-algebra of all bounded operators on the Hilbert space $\mathcal{H}_\Omega$, that will be denoted in what follows by $\mathcal{A}_\infty$.

It is worth to point it out that the set of primary transitions $\alpha_n$ generating the graph $A_\infty$ contain all the relevant information of the system.  Any transition $\alpha_{nm}$ can be obtained composing elementary transitions: $\alpha_{nm} = \alpha_{n-1}\alpha_{n+1}\cdots \alpha_m$ ($n > m$, similarly if $n < m$).

\subsubsection{The standard harmonic oscillator} 

We may define the functions $a$ and $a^\dagger$ on $\mathbf{A}_\infty$ as:
\begin{equation}\label{eq:aadagger}
a (\alpha_n^{-1}) = \sqrt{n+1} \, , \qquad a^* (\alpha_n) = \sqrt{n+1} \, , 
\end{equation}
or, alternatively, $a$ and $a^*$ are given as the formal series:
$$
a = \sum_{n = 0}^\infty \sqrt{n+1} \, \alpha^{-1}_n \, , \qquad a^* = \sum_{n = 0}^\infty \sqrt{n+1}\,  \alpha_n \, .
$$
Strictly speaking $a$, $a^*$ are not elements of the $C^*$-algebra $\mathcal{A}_\infty$, but are just functions on $\mathbf{A}_\infty$. Indeed, they define unbounded operators with a dense domain in the fundamental representation, that is in the Hilbert space $\mathcal{H}_\Omega = l^2(\mathbb{Z})$, the operators being denoted by $\mathbf{a}^\dagger = \pi_0(a^*)$ and $ \mathbf{a} = \pi_0(a)$, such that $\pi_0 (a)^\dagger = \pi_0(a^*)$. 

Moreover, as functions on $\mathbf{A}_\infty$ we can manipulate them and
a simple computation shows that:
$$
[a,a^*] = a \star a^* - a^* \star a =  \mathbf{1} \, ,
$$
with $\mathbf{1} = \sum_{n= 0}^\infty 1_n$ the unit element in $\mathcal{A}_\infty$ (note that $\pi_0(\mathbf{1}) = I$, the identity operator in $\mathcal{H}_\Omega$). Then we get the standard commutation relations for the creation and annihilation operators:
$$
[ \mathbf{a} ,  \mathbf{a}^\dagger ] = I \, .
$$
Hence we may define the Hamiltonian function:
$$
h = a^* \star a + f a^* + \bar{f} a +  \beta  =\sum_{n= 0}^\infty  n \, \delta_n  +  \sqrt{n+1} (f \alpha_n + \bar{f}\alpha_n^{-1}) +  \beta \, ,
$$
with $\omega$, $\beta$, real numbers and $f$ complex, which is the most general form of a Hamiltonian function preserving coherent states \cite{Me67} (see below).  The corresponding equations of motion are given by:
$$
\dot{a} = i[a,h] = -i  a - i f \, , \qquad \dot{a}^* =  i[a^* ,h] = i a^* + i\bar{f}  \, ,
$$
In particular, when $f = 0$, $\beta = 1/2$, we get the Hamiltonian:
$$
h_0 = \omega a^*\star  a + \frac{1}{2} = \sum_{n= 0}^\infty n \delta_n + \frac{1}{2} \, ,
$$
which constitutes the standard harmonic oscillator but written in the abstract setting of the groupoid $\mathbf{A}_\infty$, with equations of motion:
$$
\dot{a} = i[a,h] = -ia \, , \qquad \dot{a}^* =  i[a^* ,h] = ia^*  \, ,
$$
If we use the fundamental representation $\pi_0$, the Hamiltonian operator $H_0 = \pi_0 (h_0)$ may be identified with the Hamiltonian operator of a harmonic oscillator with creation and annihilation operators $\mathbf{a}^\dagger = \pi_0(a^*)$ and $ \mathbf{a} = \pi_0(a)$ respectively.

In addition to the creation and annihilation functions $a$, $a^*$ we may define the corresponding position and momentum functions $q$ and $p$ on $\mathbf{A}_\infty$ as:
$$
q = \frac{1}{\sqrt{2}} (a + a^*) \,, \qquad p = \frac{i}{\sqrt{2}} (a - a^*) 
$$
with commutation relations $[q,p] = i \mathbf{1}$.
Then the canonical Hamiltonian becomes $h_0 = (p^2 + q^2) /2$.
It is interesting to observe that, by means of the fundamental representation, the groupoid functions $q,p$ become the standard position and momentum operators $\mathbf{q} = \pi_0(q)$, $\mathbf{p} = \pi_0(p)$, which are affiliated to the $C^*$-algebra of the groupoid $\mathcal{A}_\infty$.   


\subsubsection{Coherent states} \label{sec:coherent_harmonic}

The functions $a,a^*, \mathbf{1}$, generate a three-dimensional real Lie algebra $\mathfrak{w}$ 
with commutation relations:
\begin{equation}\label{eq:comm_rel}
[a,a^*] = \mathbf{1} \, , \qquad [a,\mathbf{1} ] = [a^*, \mathbf{1}] = 0 \, ,
\end{equation}
and whose elements can be written as $\xi = \nu \mathbf{1} + i(\bar{z} a - z a^*)$, $\nu$ is a real number and $z \in \mathbb{C}$, i.e., the functions $\xi$ are real: $\xi^* = \xi$.  The Weyl-Heisenberg group $W$ (or special nilpotent group) is the unique connected and simply connected Lie group whose Lie algebra is the algebra $\mathfrak{w}$, and can be constructed by exponentiating the algebra $\mathfrak{w}$ inside the algebra $\mathcal{A}_\infty$.  

Since the operator $\pi_0(i(\bar{z} a - z a^*)) = i (\bar{z}\mathbf{a} - z \mathbf{a}^\dagger)$ is a densely defined self-adoint operator on $\mathcal{H}_\Omega$, one can construct the family of unitary operators $D(z) = e^{ z \mathbf{a}^\dagger - \bar{z}\mathbf{a} }$, and the elements of the Heisenberg-Weyl group have the form:
$$
g(\xi) =  \exp (\nu \mathbf{1} +i( \bar{z} a) - z a^*)  = e^{i\nu} e^{z \mathbf{a}^\dagger - \bar{z} \mathbf{a}}  =  e^{i\nu} D(z) \, .
 $$
(Note that because $g(\xi)$ are unitary operators they belong to $\mathcal{A}_\infty$). Then clearly $g(\xi)^{-1} = g(-\xi)$ or, denoting the Lie algebra element $\xi$ as $(\nu, z)$, we have $g(\nu, z) = e^{i\nu} D(z)$.
Hence the Weyl-Heisenberg group $W$ can be considered as a subgroup of the group of unitary elements of the $C^*$-algebra $\mathcal{A}_\infty$.

Using the BCH formula and the commutation relations (\ref{eq:comm_rel}) we get:
$$
D(z)D(w) = e^{\mathrm{Im\, } z\bar{w}/2} D(z + w) \, .
$$
Hence, $g(\nu,z) g(\nu', z') = e^{i(\nu + \nu' + \mathrm{Im\, } z\bar{w}/2)}D(z+ z') = g(\nu + \nu' + \mathrm{Im\, } z\bar{w}/2, z + z')$, that shows that the map $g \colon \mathbb{R}^3 \to W$ given by $g \colon (\nu,x,y)  \mapsto g(\nu, z)$, $z = x + iy$, defines a group isomorphism between the Heisenberg-Weyl group $W$ and the group structure defined on $\mathbb{R}^3$ by the composition law $(\nu, x,y) \circ (\nu', x',y') = (\nu + \nu' + \frac{1}{2}(xy' - yx'), x+ x', y+ y')$.

The fundamental representation of the algebra $\mathcal{A}_\infty$ provides an irreducible ryepresentation of the Weyl-Heisenberg group and a projective representation of the corresponding Abelian group $\mathbb{R}^2 \cong \mathbb{C}$ (see \cite{Ib09} and references therein).   
The Weyl-Heisenberg group $W$ has then a tautological irreducible unitary representation given by $U(g(\xi)) =e^{i\nu} e^{z \mathbf{a}^\dagger - \bar{z} \mathbf{a}} $. 
 
 If we consider an arbitrary vector $|\psi_0 \rangle \in \mathcal{H}$, and we denote by $\Gamma$ its stationary group (in the space of rays) we have that $\Gamma$ is the subgroup of elements of the form $\exp i\nu \mathbf{1}$, in which case the quotient space $M = W/\Gamma$ is the complex space $\mathbb{C}$.   The system of generalized coherent states associated to the Heisenberg-Weyl group is the set of vectors:
$$
|z \rangle = D(z) |\psi_0 \rangle \, ,
$$
that obviously satisfy $\mathbf{a} |z \rangle = z |z\rangle $
and we obtain immediately the main identity:
$$
\frac{1}{\pi} \int dz | z \rangle \langle z | = I \, .
$$
Note that in this case the fundamental representation $\pi_0$ of the groupoid and the tautological representation of the Heisenberg-Weyl group obtained from it, are both irreducible and have the same support space.  The states obtained by using both are quite difrerent.  Acting on the state $|0\rangle$ with the Heisenberg-Weyl group we obtain the standard coherent states $|z\rangle$, however acting on it with groupoid elements we obtain the vectors $|n\rangle$, i.e., the standard orthonormal basis of the Hilbert space $\mathcal{H}_\Omega$.  



\section{Generalized coherent states and groupoids}

In the previous section we have shown that the groupoids formulation of quantum mechanical systems provides a natural setting to deal with standard coherent states for the quantum harmonic oscillator.    In this section we will push this idea further to show that the groupoid formulation allows to extend the construction of coherent states in a rather flexible way.

\subsection{Generalized coherent states for groupoids}

Let $\mathbf{G}\rightrightarrows \Omega$ be a groupoid and let $(U, \mathcal{H})$ be an irreducible unitary representation with support on the total Hilbert space $\mathcal{H}$.  As before a vector on this space is denoted by the symbol $|\psi \rangle$.   Further assumptions on the properties of the representation will be introduced as needed.  

Given a vector $| \psi_0 \rangle \in \mathcal{H}$ we can extend the construction of coherent states sketched in Sect. \ref{sec:coherent_harmonic} by taking advantage of the 
canonical structures associated with the groupoid $\mathbf{G}$.  In particular we will consider the unital $C^*$-algebra $C^*(\mathbf{G})$ associated with the groupoid and its Banach-Lie group of invertible elements, that is $\gapp = \{ a \in C^*(\mathbf{G}) \mid \exists a^{-1}\, , a^{-1}a = a a^{-1} = \mathbf{1} \}$ (see \cite{Ci19} and references therein for details on Banach-Lie groups).     

The unitary representation $U$ of the groupoid $\mathbf{G}$ extends to a representation $\pi$ of the algebra $C^*(\mathbf{G})$ in $\mathcal{H}$ or, in other words, it defines a $C^*(\mathbf{G})$-module structure on $\mathcal{H}$, $a \cdot |\psi \rangle = \pi(a)|\psi \rangle$.   The representation $\pi$ defines by restriction a representation, also denoted by $\pi$, of the group $\gapp$.  We will denote by $|\psi_a \rangle$ the vector $\pi(a) |\psi_0 \rangle$.   Note that for any algebraic element $a = \sum_\alpha a_\alpha \, \alpha$ in $\gapp$, $\pi (a) = \sum_\alpha a_\alpha U(\alpha)$.   Then the definition of the $C^*$-algebra $C^*(\mathbf{G})$ as the closure in the strong topology of the algebra $\mathbb{C}[\mathbf{G}]$ implies that the representation $\pi$ of $\gapp$ is strongly continuous.  Moreover, if $U$ is irreducible, then $\pi$ is irreducible too, and in consequence the restriction to $\gapp$ is irreducible.  This implies that the family of vectors $|\psi_a\rangle$, $a\in \gapp$ spans $\mathcal{H}$.

Consider now the subgroup $\gapp_0 \subset \gapp$ that leaves invariant the state associated with the vector $|\psi_0\rangle$, that is the set of elements $c \in \gapp$ such that $\pi (c) |\psi_0 \rangle = e^{i s(c)} |\psi_0\rangle$.  Clearly $\gapp_0$ is a closed subgroup of $\gapp$ and the map $\chi \colon \gapp_0 \to \mathbb{C}^*$ given by $\chi (c) = e^{i s(c)}$ defines a one-dimensional representation of $\gapp_0$.  

We may define its homogeneous space $\mappa = \gapp /\gapp_0$ which has the structure of a smooth Banach manifold \cite{Ci19}.  We will denote the cosets in $\mappa$ as $x_a = a \gapp_0$, thus the canonical projection $\gapp \to \mappa = \gapp / \gapp_0$ reads as $a \mapsto x_a$.     We will consider now a section $\sigma$ of the canonical projection $\gapp \to \mappa$, that is, $x_{\sigma(x) } = x$.   Given any $a \in \gapp$, there is a unique element $c(a) \in \gapp_0$ that depends on the section $\sigma$ such that $a = \sigma(x_a) c(a)$.  Clearly we have, 
$$
|\psi_a \rangle = \pi(a) |\psi_0 \rangle = \pi(\sigma(x_a)) \pi (c(a)) |\psi_0\rangle = e^{s(a)} |\psi_{\sigma(x_a)} \rangle \, ,
$$ 
with $s(a) = s (c(a))$ by definition.  If we just denote by $|x \rangle$ the vector $\pi(\sigma(x)) |\psi_0 \rangle = |\psi_{\sigma(x)} \rangle$, we get 
\begin{equation}\label{eq:xpsia}
|x \rangle = e^{-i s(a)} |\psi_a \rangle
\end{equation}
for any $a \in \gapp$ in the same coset as $\sigma (x)$, that is such that $x_a = x$.    In general it is not possible to guarantee that the family of vectors $|x \rangle$ will span $\mathcal{H}$ even if $\pi$ is irreducible.

Again a simple computation shows that $\sigma (a c) = \sigma (a ) + \sigma (c)$, $a \in \gapp$, $c \in \gapp_0$ and $\pi(a') |x \rangle = e^{i(\sigma (a'a) - \sigma(a))} |a' x \rangle$, with $a$ in the coset $x$ and $a' x$ denotes the natural action of $\gapp$ on $\mappa$:
\begin{eqnarray}\label{eq:aprimex}
\pi(a') |x \rangle &=&  e^{-i\sigma(a)}\pi(a') \pi(a) |\psi_0 \rangle =   e^{-i\sigma(a)}\pi(a'a) |\psi_0 \rangle  \\ &=&  e^{-i\sigma(a)} |\psi_{a'a} \rangle =   e^{-i\sigma(a)}e^{i\sigma(a'a)} |x_{a'a} \rangle =   e^{i(\sigma(a'a) -\sigma(a))} |a'x  \rangle \, . \nonumber
\end{eqnarray}
It is clear that the combination $\sigma(a'a) -\sigma(a)$ in the exponent in the last term in the previous equation does not depend on $a$ but just on the coset $x$. In what follows we will denote by $|0 \rangle$ the vector $|\psi_0\rangle$. 
Finally using Eq. (\ref{eq:xpsia}) we get:
\begin{equation}\label{eq:xxprime}
\langle x | x' \rangle = e^{i(\sigma(a) - \sigma(a'))} \langle 0 | \pi(a^*a') | 0 \rangle \, .
\end{equation}

The Banach-Lie group $\gapp$ is too large for many applications, in general is an infinite-dimensional Banach-Lie group and being modelled on an infinite dimensional Banach space is not locally compact.   It is not necessary to consider the orbit  $\{| \psi_a \rangle \mid a \in \gapp \}$ of the full group $\gapp$ of invertible elements to construct a set of states suitable for the description of a given dynamics.  

Given a dynamics $h$ on $\mathbf{G}$ with associated flow $\varphi_t$ of automorphisms of $C^*(\mathbf{G})$ it would be much more desirable to consider a subset $M \subset \gapp$ compatible with it, that is $\varphi_t (m) = m_t \in M$ and such that the corresponding space of states will span the Hilbert space $\mathcal{H}$.   The last condition is guaranteed if a completeness condition of the form (compare with \cite[Sect. 2]{Ci17}):
\begin{equation}\label{eq:mcomplete}
\int_M d\mu (m) |m \rangle\, \langle m | = I \, ,
\end{equation}
is satisfied, $I$ denotes the identity map in $\mathcal{H}$ and $\mu$ is a measure on $M$.  Notice that if $M$ is invariant under the dynamics of $h$, then we have:
\begin{equation}\label{eq:mstable}
U_t |m \rangle = |m_t \rangle
\end{equation}
with $U_t = e^{it H} = \pi (\exp (it h))$,  the strongly continuous one-parameter group defined by the dynamics $h$ with Hamiltonian $H$.   

Thus in many occasions it would suffice to consider a finite-dimensional subgroup $G \subset \gapp$ of unitary elements, as we did for instance in the discussion of standard coherent states for the harmonic oscillator, Sect. \ref{sec:coherent_harmonic}.  In that ocassion we considered the Heisenberg-Weyl group $W \subset \gapp (\mathcal{H})$, a subgroup of the group $\gapp (\mathcal{H})$ of bounded invertible operators in $\mathcal{H}$, defined by the exponentiation of the Lie algebra $\mathfrak{w}$ in $\mathcal{A}_\infty$.  Again the restriction of the representation $\pi$ to $G$ will define a strongly continuous representation of $G$ and the previous discussion applies \textit{mutatis mutandis} except for the fact that even if $\pi$ is irreducible, the restricted representation can be reducible. This did not happen with the Heisenberg-Weyl group but it would have been the situation related to the group $SU(2)$ using the Jordan-Schwinger map to construct a subgroup of $\gapp(\mathcal{H})$.     

In what follows we  will assume that we are considering a finite-dimensional locally compact subgroup $G$ of unitary elements in $\gapp$ and we will restrict ourselves to an irreducible subrepresentation of the representation defined on it by the restriction of the representation $\pi$ of $\gapp$ that will be denoted without risk of confusion with the same notation $(\pi, \mathcal{H})$.  It will also be assumed that the representation is square integrable, then we will denote as before by $|m\rangle $ the vector $\pi(\sigma (m_g))|0\rangle$. It is associated with the choice of a section $\sigma$ of the canonical projection $\rho \colon G \to M$, where $M = G/G_0$ is the homogeneous space consisting of the cosets of the isotropy group of the state associated to the vector $|0\rangle$.  The representation $\pi$ being square integrable implies that $\int_G \pi(g) |\psi \rangle d\mu_g(g) \in \mathcal{H}$ (at least for a dense subspace of $\mathcal{H}$).

We will also assume for simplicity that the group $G$ is unimodular, that is the canonical left-invariant Haar measure $\mu_G$ on $G$ is right-invariant. Therefore one can induce an invariant measure, also denoted by $\mu$, on the homogeneous space $M$, provided that the volume of $G_0$ is finite, as:
\begin{equation}\label{eq:muG}
\mu (\Delta ) = \frac{1}{\mu_G(G_0)} \mu_G(\rho^{-1}(\Delta)) \, ,
\end{equation}
and $\Delta$ is measurable if $\rho^{-1}(\Delta)$ is measurable.  Equations (\ref{eq:xpsia}), (\ref{eq:aprimex}) and (\ref{eq:xxprime}), become:
\begin{equation}\label{eq:mpsia}
|m \rangle = e^{-i s(g)} |\psi_g \rangle \, ,
\end{equation}
\begin{equation}\label{eq:aprimem}
\pi(g') |m \rangle = e^{i(\sigma (g'g) - \sigma(g))} |g' m \rangle \, ,
\end{equation}
and
\begin{equation}\label{eq:mmprime}
\langle m | m' \rangle = e^{i(\sigma(g) - \sigma(g'))} \langle 0 | \pi(g^{-1}g') | 0 \rangle \, .
\end{equation}

Under such conditions the completeness of the familly of coherent states $|m \rangle$ can be assessed as follows.
Consider the operator:
$$
C = \int d\mu (m) |m \rangle \langle m | \, .
$$
The operator $C$ is well defined because the measure $\mu_G$ is right-invariant and the subgroup $G_0$ has finite volume. Indeed, using (\ref{eq:mpsia}) and (\ref{eq:muG}), we have:
\begin{eqnarray*}
C |\psi \rangle &=& \int d\mu (m) |m \rangle \langle m | \psi \rangle =  \int d\mu (m) \langle m | \psi \rangle  e^{-i s(g)} |\psi_g \rangle \\ & =& \frac{1}{\mu_G(G_0)} \int_G d\mu_G (g)  \langle m |\psi \rangle  e^{-i s(g)} \pi(g) |\psi_0\rangle \, ,
\end{eqnarray*}
which is well-defined because the representation is square integrable.
Now it is clear that $C$ commutes with the irreducible representation $\pi$ of the group $G$. Using (\ref{eq:aprimem}), in fact, we get:
\begin{eqnarray*}
\pi (g) C \pi(g)^\dagger &=&  \int d\mu (m) \pi (g)|m \rangle \langle m | \pi(g)^\dagger = \int d\mu (m) |gm \rangle \langle gm | \\ & =& \frac{1}{\mu_G(G_0)} \int_G d\mu_G (g) |gm \rangle \langle gm |  =  \int d\mu (m) |m \rangle \langle m | = C \, .
\end{eqnarray*}

Because of Schur's Lemma, $C$ must be a multiple of the identity:
$$
C = \lambda I \, 
$$
We obtain the factor $\lambda$ by computing the expected value of $C$:
$$
\lambda = \langle 0 | C |0 \rangle = \int_M |\langle 0 | m \rangle |^2 d\mu (m)  \, ,
$$
and the family of projectors $P_m$ associated to the generalized coherent states $|m \rangle$ provides a resolution of the identity:
$$
\frac{1}{\lambda} \int_M P_m d\mu (m) = I \, .
$$ 
Moreover, we get 
$$
|\psi \rangle = \frac{1}{\lambda} \int_M \langle m | \psi \rangle  | m \rangle d\mu (m) \, ,
$$
and the function 
$$
k(m,m') = \frac{1}{\lambda} \langle m | m' \rangle = \frac{e^{i(\sigma(g) - \sigma(g'))}}{\lambda} \langle 0 | \pi(g^{-1}g') | 0 \rangle \, ,
$$ 
defines a reproducing kernel Hilbert space structure on the space of function $\mathcal{F}(M) = \{ f\colon M \to \mathbb{C}  \mid f(m ) = \langle \psi | m \rangle\, , | \psi \rangle \in \mathcal{H}  \}$.  In fact:
$$
k(m,m') = \frac{1}{\lambda} \int_\Omega k(m,m'') k(m'',m') d \mu (m'') \, .
$$
and the inner product on the space $\mathcal{F}(\Omega) $ is given by:
$$
\langle f | g \rangle = \int_\Omega \overline{f(m)} k(m,m') g(m') d\mu (m) d\mu(m') \, .
$$

\medskip



\subsection{$f$-oscillators and other generalized coherent states}

An immediate generalization of the results presented in section \ref{harmonic oscillator} is provided by the so-called f-oscillators. They were introduced in the \cite{Ma97} in order to study the effect of non-linear non-canonical trasformations which preserve the dynamics, both at the classical and the quantum level. In this subsection we will describe this dynamical system according to the groupoid formulation outlined before. 

As already explained in Section \ref{harmonic oscillator}, the starting point is the space of possible physical outcomes and transitions, which provides the kinematical background in the groupoid picture. Similarly to the usual harmonic oscillator, f-oscillators have a discrete countable set of outcomes, $a_n$ with $n \in \mathbb{N}$, with morphisms generated by the transitions $\alpha_n  :  a_n  \rightarrow  a_{n+1}$. Therefore, the whole family of harmonic and f-oscillators share the common kinematical background, which is the groupoid $\mathbf{A}_{\infty}$. Analogously, it is possible to build the $C^*$-algebra, $\mathcal{A}_{\infty}$ associated with the groupoid $\mathbf{A}_{\infty}$ and again this coincides with the unital $C^*$-algebra of all bounded operator over the Hilbert space $\mathcal{H}_{\Omega}=\ell^2(\mathbb{Z})$. The difference among the different systems of this family is given by the dynamics, which amounts to assign specific values to objects and transitions of the groupoids. In particular for a f-oscillator we can define the following functions $A_{f}$ and $A_f^{*}$ (compare with Eq. (\ref{eq:aadagger})):
\begin{equation}
A_f(\alpha_n^{-1})=f(n+1)\sqrt{n+1}\,, \qquad A_f^*(\alpha_n)=f(n+1)\sqrt{n+1}\,,
\end{equation} 
which, on the fundamental representation, will define operators whose domains will depend on the chosen function $f(n)$. Their commutator is the function $\mathbf{F}$ which can be expressed as
\begin{equation}
\mathbf{F} = \left[ A_f , A^*_f \right] = A_f \star A^*_f - A^*_f \star  A_f  = \sum_{n=0}^{\infty}\left( f^2(n+1)(n+1) + f^2(n)n \right)\delta_n \, .
\end{equation}
The Hamiltonian function, instead, is defined as the function 
\begin{equation}
h_f = \frac{\omega}{2} \left( A_f \star A^*_f + A^*_f \star A_f \right) = \sum_{n=0}^{\infty}\frac{\omega}{2}\left( f^2(n+1)(n+1) + f^2(n)n \right) \delta_n\,.
\label{f-oscillator_hamiltonian}
\end{equation} 
In the fundamental representation $\pi_0$ we can associate to these functions the operators 
$$
\mathbf{A}_f = \pi_0(a_f)=\mathbf{a}f(\mathbf{n})\, ,\qquad  \mathbf{A}_f^{\dagger}=\pi_0(A^*_f) = f(\mathbf{n})\mathbf{a}^{\dagger} \, , \quad H_f=\pi_0(h_f)\, ,
$$ 
where $\mathbf{n}=\mathbf{a}^{\dagger}\mathbf{a}$. The evolution of the operator $\mathbf{A}_f$ is given in terms of the one parameter group of unitary transformations $U(t)= e^{-itH_f}$ as follows:
$$
\mathbf{A}_f(t) = U^{\dagger}(t)\mathbf{a}f(\mathbf{n})U(t)= \mathbf{a}f(\mathbf{n})e^{-i\omega(\mathbf{n})t}\,, 
$$
where $\omega(\mathbf{n}) = \frac{1}{2}\left( (\mathbf{n}+1)f^2(\mathbf{n}+1) -\mathbf{n}f(\mathbf{n}) \right)$.

As already explained in the previous part of this section, it is possible to use unitary operators in $\gapp$ to construct families of states of the Hilbert space $\mathcal{H}_{\Omega}$. Even if we focused the properties connected with families of states which are obtained by means of a unitary representation of some Lie group $G$, one can consider also other subsets of states $\{ | m\rangle \}$ satisfying the completeness and stability conditions as described by Eqs. (\ref{eq:mcomplete}), (\ref{eq:mstable}). 

In \cite{Ma97} non-linear f-coherent states were introduced as eigenstates of the $\mathbf{A}_f$ operator, having the following expression:
\begin{equation}
|z,f \rangle = N_{f,z}\sum_{n=0}^{\infty}\frac{z^n}{\sqrt{n!}[f(n)]!}|n\rangle\,.
\end{equation}
It is possible to define these states on subsets of the complex plane and for suitable functions $f$, such that the normalization factor $N_{f,z}$ is finite. Moreover, if the function $f$ is also invertible, a completeness relation can be proven to hold.

Following an analogous procedure to the one used to define coherent states for the harmonic oscillator, we will consider the strongly continuous map, which associates to each complex number $z \in \mathbb{C}$ the unitary operator $D_f(z)$ defined as follows:
\begin{equation}
D_f(z) = e^{zA_f - \bar{z}A^*_f}\,.
\end{equation}  
Contrarily to the harmonic oscillator case, these operator do not form a projective representation of the additive group of complex numbers. However, one can define the set of states $\left\lbrace |z_f\rangle \right\rbrace$ acting on the vacuum state $|0\rangle$:
\begin{equation}
|z_f\rangle = D_f(z)|0\rangle\,.
\end{equation}
If the function $f$ is decreasing, we can write the operator $D_f(z)$ as a series expansion and one gets the following expression for this set of generalized coherent states:
$$
|z_f\rangle = \sum_{n=0}^{\infty}\sum_{k=0}^n (-1)^k |z|^{2k}  \frac{\sqrt{(2(n-k))!}}{(2n)!}   \Big( \beta_{n,2k} z^{2(n-k)} \left[ f(2(n-k)) \right]! |2(n-k)\rangle $$ 
\begin{equation*}
- \frac{ \sqrt{2(n-k)+1}}{2n+1} \beta_{n,2k+1} z^{2(n-k)+1}\left[ f(2(n-k)+1) \right]! |2(n-k)+1\rangle \Big) \,,
\end{equation*}
where $\left[f(k)\right]!=f(k)f(k-1)\cdots f(1)$ and the coefficients $\beta_{i,j}$ are functions of the integer variables $(n,k)$. In particular $\beta_{n,0} = \beta_{n,1} = 1$, whereas the coefficients $\beta_{n,2k}$ and $\beta_{n,2k+1}$, with $k>1$, can be written in therms of $k$ indices $j_i$, $1\leq i \leq k$, as follows:
$$
\beta_{n,2k} = \sum_{j_1=0}^{2(n-k)}\sum_{j_2=0}^{j_1+1}\cdots\sum_{j_k=0}^{j_{k-1}+1}G_{j_1}(1)G_{j_2}(3)\cdots G_{j_k}(2k-1)
$$
$$
 \beta_{n,2k+1} = \sum_{j_1=0}^{2(n-k)+1}\sum_{j_2=0}^{j_1+1}\cdots\sum_{j_k=0}^{j_{k-1}+1}G_{j_1}(2)G_{j_2}(4)\cdots G_{j_k}(2k)\,,
$$
where $G_k(n)=\sum_{j=0}^{k}\left[ \mathbf{F}(n+j) \right]$. The evolution of these states in the Schr\"{o}dinger picture can be straightforwardly computed, consisting in the multiplication by the time dependent phase factor $e^{-iE(n)t}$ of any vector $|n\rangle$, and one can immediately notice that only the hamiltonian function $h_0$ of the harmonic oscillator maps coherent states into coherent states. 

Moreover, since the set $\left\lbrace D_f(z) \mid z\in \mathbf{C}\right\rbrace$ does not form an irreducible unitary representation  of the Heisenberg-Weyl group, its completeness must be assessed differently and a case-by-careful analysis  depending on the properties of the function $f$ should be performed \cite{Ma97}.  





\section{Conclusions and discussion}
It has been shown that the groupoid picture of Quantum Mechanics provides a natural background to construct families of generalized coherent states.   The fundamental representation of the groupoid describing the given quantum system together with a sensible choice of a subgroup of the group of unitary elements in the $C^*$-algebra of the groupoid compatible with a given dynamics will provide one such family.    In this sense the classical construction of coherent states from unitary group representations fits naturally in this description.    Beyond this, any other choice of a subset of elements in the $C^*$-algebra of the groupoid satifying natural invariance and completeness conditions will suffice to define generalized coherent states.  This idea is illustrated by deforming the harmonic oscillator dynamics in the groupoid picture to provide families of f-coherent states.

The completeness property of such generalized families can be characterized by using the theory of frames (see for instance \cite{An08} an references therein), a task that will be considered elsewhere.  


\section*{Acknowledgements}
A.I. and G.M. acknowledge financial support from the Spanish Ministry of Economy and Competitiveness, through the Severo Ochoa Programme for Centres of Excellence in RD (SEV-2015/0554).
A.I. would like to thank partial support provided by the MINECO research project MTM2017-84098-P and QUITEMAD++, S2018/TCS-A4342.
G.M. would like to thank the support provided by the Santander/UC3M Excellence Chair Programme 2019/2020.


\end{document}